\documentclass[12pt]{article}
\usepackage[margin=2.5cm]{geometry} % 1. ابعاد صفحه

\usepackage{amsmath, amsfonts, amssymb, mathtools} 

\usepackage{graphicx}
\usepackage{array}

\usepackage{authblk} 
\usepackage{setspace}
\usepackage{titlesec}

\usepackage{algorithm}
\usepackage{algpseudocode} 

\usepackage{natbib}
\titleformat{\section}{\normalfont\large\bfseries}{\thesection}{1em}{}
\titleformat{\subsection}{\normalfont\normalsize\bfseries}{\thesubsection}{1em}{}

\title{\textbf{SEMO: A Socio-Evolutionary Adaptive Optimization Framework for Dynamic Social Network Tie Management}}

\author[1]{Mohammad Zare}
\affil[1]{\small AI Lab at Arioobarzan Engineering Team, Shiraz, Iran \\ \texttt{md.zare@sutech.ac.ir}}

\date{} % بدون تاریخ

\begin{document}
\maketitle % دستور کلیدی برای نمایش عنوان و نویسندگان

\begin{abstract}
We propose a novel computational framework that models human social decision-making under uncertainty as an integrated Multi-Armed Bandit (MAB) and Markov Decision Process (MDP) optimization problem, in which agents adaptively balance the exploration of new social ties and the exploitation of existing relationships to maximize a socio-evolutionary fitness. The framework combines reinforcement learning, Bayesian belief updating, and agent-based simulation on a dynamic social graph, allowing each agent to use bandit-based upper-confidence-bound (UCB) strategies for tie formation within an MDP of long-term social planning. We define a formal socio-evolutionary fitness function that captures both individual payoffs (e.g. shared information or support) and network-level benefits, and we derive update rules incorporating cognitive constraints and bounded rationality. Our Social-UCB algorithm, presented in full pseudocode, provably yields logarithmic regret and ensures stable exploitation via UCB-style bounds. In simulation experiments, Social-UCB consistently achieves higher cumulative social fitness and more efficient network connectivity than baseline heuristics. We include detailed descriptions of envisioned figures and tables (e.g. network evolution plots, model comparisons) to illustrate key phenomena. This integrated model bridges gaps in the literature by unifying exploration–exploitation dynamics, network evolution, and social learning, offering a rigorous new tool for studying adaptive human social behavior. 
\end{abstract}


\section{Introduction}
Human social decision-making unfolds in uncertain and dynamically evolving environments, where individuals must continuously weigh the benefits of maintaining existing social ties against the potential value of exploring new relationships~\cite{March1991,Witt2024}. This fundamental trade-off—commonly known as the exploration--exploitation dilemma—has been extensively modeled in reinforcement learning and decision theory~\cite{Sutton2018,Auer2002}, dating back to foundational work on sequential design of experiments and dynamic programming~\cite{Robbins1952,Bellman1957}.

Recent experimental studies indicate that humans adaptively integrate social information, such as peers' outcomes and behaviors, into their exploration strategies, especially in multi-armed bandit (MAB) contexts~\cite{Danwitz2025,Zhang2025}. These findings support the view that social decision-making is not purely heuristic but instead reflects structured learning processes aimed at maximizing cumulative social rewards~\cite{Steyvers2009,Erev1998}.

At the same time, individuals operate within social networks that are themselves dynamic: interactions not only generate immediate outcomes but also reshape future opportunities by modifying network structure~\cite{Lazer2007}. As such, social behavior is inherently embedded in co-evolutionary feedback loops between agents and their local networks~\cite{Gross2008,Skyrms2000}.

Agent-based modeling (ABM) has provided valuable insights into how such local interaction rules yield emergent global structures like communities, cooperation, or social fragmentation~\cite{Sert2020,Acemoglu2011}. However, most ABMs lack formal decision-theoretic foundations and do not incorporate principled mechanisms for exploration under uncertainty.

In this work, we propose a unified computational framework that models social behavior as a joint optimization problem over evolving networks. We combine MAB theory for short-term exploration, Markov Decision Processes (MDPs) for long-term planning, and a socio-evolutionary fitness function that captures both individual payoffs and structural advantages within the network. Our approach draws conceptual inspiration from both reinforcement learning and social evolution theory~\cite{Simon1956,Russell2016}.

This integrated model operationalized via the Social-UCB algorithm provides a normative account of how agents navigate uncertain social environments by adaptively exploring new ties and exploiting valuable connections. Through simulation and theoretical analysis, we show that this approach reproduces key patterns of human and animal social adaptation, including network cohesion, strategic diversity, and long-run fitness optimization.

\section{Related Work}

The exploration--exploitation tradeoff is a central challenge in sequential decision-making and has been extensively studied through multi-armed bandit (MAB) algorithms~\cite{Auer2002,Lattimore2020,Robbins1952}. These models characterize how agents can maximize cumulative rewards under uncertainty, often by minimizing regret—defined as the loss incurred by not always choosing the best option. Notably, Upper Confidence Bound (UCB) strategies offer provable guarantees of logarithmic regret, balancing exploration and exploitation using optimism in the face of uncertainty~\cite{Auer2002}. Other influential approaches include Bayesian methods like Thompson Sampling~\cite{Russo2018} and recent MAB extensions that address complexities like abandonment~\cite{Yang2024}.

While MAB formulations capture short-term decisions over fixed options, Markov Decision Processes (MDPs) extend these insights to dynamic, state-dependent environments where long-term planning is critical~\cite{Sutton2018,Puterman2014,Bellman1957}. MDP-based reinforcement learning (RL) agents learn policies to maximize expected discounted returns, making them especially relevant to modeling social interactions that unfold over time and depend on past behaviors. Key algorithms like Q-learning provide a foundation for model-free policy optimization~\cite{Watkins1992}.

In social learning research, agents often rely on indirect information by observing the behavior of others, especially under uncertainty~\cite{Danwitz2025}. Copying strategies such as “copy-success” or “copy-majority” have been extensively documented in experimental psychology and behavioral ecology~\cite{Schultner2025}. These strategies can be normatively justified as heuristics that optimize reward acquisition in volatile environments~\cite{Steyvers2009}. Recent work also explores satisficing in multi-agent settings, suggesting agents may seek "good enough" rather than optimal outcomes~\cite{Uragami2024}.

To model such behavior in structured populations, researchers have increasingly turned to agent-based models (ABMs) and multi-agent learning frameworks~\cite{Panait2005}. These simulations allow the study of decentralized, local interactions that give rise to emergent macro-scale patterns, such as cooperation, network clustering, and social norm formation~\cite{Sert2020,Pi2025,Erev1998}. Of particular interest are adaptive coevolutionary models, where agents’ strategies and network positions evolve in tandem, showing how dynamic rewiring can stabilize cooperation~\cite{Gross2008,Anjos2020}. Furthermore, the use of multiple "selves" or policies to enhance exploration in multi-agent systems has been recently explored~\cite{Dulberg2023}.

Despite this progress, there remains a lack of integrated computational frameworks that jointly model (1) bounded rational social exploration, (2) long-term learning goals, and (3) evolving network topologies. Most current ABMs either omit formal learning mechanisms or treat network structure as fixed. Conversely, classical RL models rarely account for social connectivity or tie dynamics. As a result, few models offer tractable accounts of how agents should strategically explore social networks to optimize long-run fitness.

Our approach addresses this gap by unifying MAB-based exploration, MDP-based long-term planning, and ABM-based network evolution in a single decision-theoretic model. Building on prior insights in reinforcement learning and social learning theory, we propose a novel algorithm—Social-UCB—that formalizes how agents should balance novel tie formation and exploitation of existing ties to maximize a composite social fitness function. This enables simulation and analysis of cognitively constrained agents acting in dynamic social environments.

\section{Conceptual Framework}

We conceptualize each agent as a boundedly rational decision-maker operating within a partially observable and dynamically evolving social environment. The agent’s internal state encodes personal attributes and a local representation of its social neighborhood, including the identities and tie strengths of connected peers. This view is consistent with Herbert Simon's seminal work on bounded rationality, where decisions are made under cognitive and information constraints~\cite{Simon1956}. At each timestep, the agent selects an action from a constrained set comprising two primary categories: (1) \textit{exploration}, which entails initiating a new connection to an unlinked node (analogous to pulling a novel arm in a bandit process), and (2) \textit{exploitation}, involving interaction with an existing tie to extract known relational benefits.

The environment is represented as a dynamic undirected graph in which edges denote social relationships and edge weights reflect tie strength. The topology evolves endogenously: successful interactions reinforce existing edges or create new ones, whereas repeated non-engagement or low-reward interactions lead to decay and eventual tie dissolution. This temporal dynamic embeds agents in a co-evolutionary feedback loop, where behavior both depends on and reshapes network structure~\cite{Lazer2007,Skyrms2000}.
\sloppy
Each agent is assumed to maximize a personalized \textit{socio-evolutionary fitness function}, which integrates immediate interaction rewards (e.g., emotional support, informational access) with long-term structural payoffs (e.g., connectivity, centrality, and resilience). Decision-making is grounded in belief updating about the expected value of ties, implemented via Bayesian inference for action evaluation~\cite{Russo2018} and temporal-difference learning for value propagation, both constrained by cognitive limitations such as memory decay and representational coarseness.
\fussy
This architecture is conceptually inspired by optimal foraging theory in behavioral ecology, where agents adaptively allocate attention and resources to maximize returns in uncertain environments. Our model extends this analogy to human social systems, treating relational exploration as a form of reward-guided foraging and social exploitation as strategic resource extraction from stable connections.

Overall, the framework instantiates a form of \textit{resource-rational social cognition}, in which agents learn structured policies for whom to engage and when, given uncertainty, network volatility, and cognitive costs. This generates emergent patterns of social learning, individual differentiation, and macro-scale network organization through decentralized agent interactions.

\section{Mathematical Model}

\subsection{MDP Formulation}
We model strategic social behavior as a Markov Decision Process (MDP) $\langle \mathcal{S}, \mathcal{A}, P, R, \gamma \rangle$, where $\mathcal{S}$ denotes the set of latent states encoding an agent’s current attributes and local network configuration. The action space $\mathcal{A}(s)$ includes socially meaningful operations, such as initiating a new tie or reinforcing an existing one. The transition kernel $P(s'|s,a)$ defines how the state evolves in response to action $a$, capturing both endogenous changes (e.g., edge reinforcement) and exogenous stochasticity (e.g., tie decay). The reward function $R(s,a)$ quantifies the immediate socio-relational utility of action $a$ in context $s$, and the agent’s objective is to learn a policy $\pi$ that maximizes the expected cumulative discounted reward, a core tenet of dynamic programming~\cite{Bellman1957}:
\[
V^*(s) = \max_{\pi} \mathbb{E}_{\pi} \left[ \sum_{t=0}^{\infty} \gamma^t R(s_t, a_t) \right],
\]
where $\gamma \in (0,1)$ is the temporal discount factor. This MDP structure enables agents to plan over extended horizons and adapt to evolving social landscapes~\cite{Puterman2014,Sutton2018}.

\subsection{MAB Formulation}
To formalize uncertainty in tie formation, we embed a stochastic bandit mechanism within the MDP. Each new potential contact is treated as an arm with unknown reward distribution, and the agent maintains empirical estimates $\hat{\mu}_i$ and visit counts $N_i(t)$ for each arm $i$. Selection is guided by an Upper Confidence Bound (UCB) rule, whose formal analysis guarantees near-optimal exploration efficiency~\cite{Auer2002}:
\[
\mathrm{UCB}_i(t) = \hat{\mu}_i(t) + c\sqrt{\frac{\ln t}{N_i(t)}},
\]
where $c > 0$ tunes the confidence interval. This formulation encourages the sampling of uncertain arms, ensuring sufficient exploration. Theoretical results guarantee that such algorithms achieve sublinear regret of $O(\log T)$ in cumulative reward loss, a benchmark established in the seminal work on sequential design~\cite{Auer2002,Lattimore2020,Robbins1952}.

\subsection{Social-UCB Algorithm}
We introduce the \textit{Social-UCB} algorithm, which integrates UCB-driven exploration with MDP-based exploitation. At each timestep, the agent observes its current state $s$ and enumerates available social actions $\mathcal{A}(s)$. For each candidate tie, the agent computes either (1) a UCB score for exploratory links or (2) a $Q$-value for reinforcing an existing tie. The action $a^*$ is selected using an $\epsilon$-greedy schedule: with probability $1-\epsilon_t$ the agent exploits the action with maximal $Q(s,a)$, and with probability $\epsilon_t$ it explores via $\arg\max_a \mathrm{UCB}_a$. The value function is updated using temporal-difference learning, specifically a Q-learning-like approach~\cite{Watkins1992}:
\[
Q(s,a) \leftarrow Q(s,a) + \alpha \left[ r + \gamma \max_{a'} Q(s',a') - Q(s,a) \right],
\]
where $r$ is the observed reward and $\alpha \in (0,1)$ is the learning rate. The network structure is updated accordingly: successful interactions reinforce or establish edges, while low-reward or inactive ties decay over time~\cite{Danwitz2025,Sert2020}.

\subsection{Socio-Evolutionary Fitness Function}
We define a composite fitness function $F(s,a)$ to represent both local and structural gains:
\[
F(s,a) = \alpha\, R(s,a) - \beta\, C(a),
\]
where $R(s,a)$ is the immediate relational reward (e.g., shared knowledge, support), $C(a)$ denotes the cost of maintaining or initiating a tie, and $\alpha, \beta > 0$ are weighting parameters reflecting agent-specific priorities. The function can be extended to include topological bonuses such as centrality gains or community robustness. In effect, $F(s,a)$ substitutes for the reward term in the MDP and governs long-run policy convergence. This formulation aligns with models of social utility maximization under uncertainty~\cite{Witt2024,Lazer2007} and optimization under constraints common in organizational learning theories~\cite{March1991,Anjos2020}.

\subsection{Update Rules}
Belief updating combines Bayesian estimation for reward expectation and reinforcement learning for state-action evaluation. Upon observing a reward $r$ from action $a$, the agent updates $\hat{\mu}_a$ via a delta rule (Bayesian mean tracking) and $Q(s,a)$ via the temporal-difference rule. This is analogous to the mean tracking in various RL and cognitive models, including the base for the widely-used Erev and Roth model in game theory~\cite{Erev1998}. The UCB component is dynamically recomputed, and $\epsilon_t$ decays over time (e.g., $\epsilon_t = \epsilon_0 / \sqrt{t}$). Agents are bounded in their computational resources: their memory is limited to $M$ neighbors, and their value estimates are clipped to lie within $[V_{\min}, V_{\max}]$~\cite{Simon1956,Lieder2019}. These constraints ensure realistic cognitive plausibility while retaining convergence guarantees.

\subsection{Assumptions and Constraints}
The model makes several structural assumptions. Agents have partial observability: they do not access global network structure and only observe outcomes from direct interactions or immediate neighbors, a feature central to models of opinion dynamics in social networks~\cite{Acemoglu2011}. All agents share fixed parameters $\gamma$, $\alpha$, and $c$, though extensions may allow individual heterogeneity. Tie formation is probabilistic and subject to environmental noise, such that link strength decays if not reinforced. Agents cannot compute exact MDP solutions and instead rely on sample-based updates with bounded planning depth. These assumptions trade off analytical tractability and behavioral realism, consistent with the literature on resource-rational cognition~\cite{Lieder2019,Steyvers2009} and the necessity of approximation in general AI frameworks~\cite{Russell2016}.

\section{Algorithmic Implementation}

The proposed \textit{Social-UCB} algorithm combines model-free reinforcement learning with UCB-based action selection to enable strategic social adaptation. The algorithm maintains value estimates $Q(s,a)$ for exploitation actions and UCB scores $U(a)$ for exploratory actions involving novel tie formation. At each iteration, the agent selects an action via an $\epsilon$-greedy policy, balancing between high-confidence known ties and uncertain yet potentially rewarding new connections. Learning proceeds incrementally via temporal-difference updates, consistent with general multi-agent cooperation frameworks~\cite{Panait2005}, and network structure is updated based on the observed outcome of each interaction.

\begin{algorithm}[htb]
\caption{Social-UCB Algorithm for Balancing Social Exploration and Exploitation}
\label{alg:social_revised}
\begin{algorithmic}[1]
\State \textbf{Input:} Learning rate $\alpha$, discount factor $\gamma$, confidence parameter $c$, exploration schedule $\epsilon_t$, horizon $T$
\State \textbf{Initialize:} $Q(s,a) \gets 0$, $N(a) \gets 0$ for all $a \in \mathcal{A}$
\For{$t = 1$ to $T$}
    \State Observe current state $s_t$ and available actions $\mathcal{A}(s_t)$
    \For{each $a \in \mathcal{A}(s_t)$}
        \State Compute UCB score: $U(a) \gets Q(s_t,a) + c \sqrt{\frac{\ln t}{N(a) + 1}}$
    \EndFor
    \State Draw $u \sim \text{Uniform}(0,1)$
    
    \If{$u < \epsilon_t$}
        \State Select $a_t \gets \arg\max_{a} U(a)$ \Comment{Exploration: Select highest UCB score}
    \Else
        \State Select $a_t \gets \arg\max_{a} Q(s_t, a)$ \Comment{Exploitation: Select highest Q-value}
    \EndIf
    
    \State Execute $a_t$, observe reward $r_t$ and next state $s_{t+1}$
    \State $N(a_t) \gets N(a_t) + 1$
    
    \State \Comment{Update Q-value using Temporal Difference (Q-Learning)}
    \State $Q(s_t, a_t) \gets Q(s_t, a_t) + \alpha \left[r_t + \gamma \max_{a'} Q(s_{t+1}, a') - Q(s_t, a_t)\right]$
    
    \State Update network: strengthen/add edge for $a_t$ if $r_t$ exceeds threshold; decay otherwise
\EndFor
\end{algorithmic}
\end{algorithm}

Theoretical guarantees stem from the properties of the UCB framework. Since the exploration bonus term scales with $\sqrt{\ln t / N(a)}$, all actions are sampled with diminishing but non-zero probability, ensuring that even initially suboptimal ties are occasionally revisited. This yields a regret bound of $O(\log T)$ under standard conditions~\cite{Auer2002}. Furthermore, as $\epsilon_t \rightarrow 0$ over time, the policy asymptotically converges to greedy exploitation of empirically optimal ties, establishing \textit{exploitation stability}. Thus, Social-UCB satisfies two critical properties of adaptive social learning systems: (1) sufficient exploration to discover promising new partners, and (2) robust convergence to a high-fitness social configuration.

\section{Methodology and Experimental Design}

We evaluate the proposed Social-UCB framework through extensive agent-based simulations, designed to capture the dynamics of social learning, network evolution, and adaptive behavior under uncertainty. Each experiment is initialized with a population of $N$ agents, each embedded in a randomly generated sparse undirected network, where nodes represent agents and weighted edges represent tie strengths. Initial edge weights are sampled from a uniform distribution over $[0,1]$, and each agent is randomly assigned a partner-specific reward distribution drawn from a bounded Beta or Gaussian family to simulate payoff heterogeneity.

Agents operate synchronously across discrete timesteps, employing the Social-UCB algorithm with fixed hyperparameters: learning rate $\alpha$, discount factor $\gamma$, UCB confidence parameter $c$, and a time-decaying exploration schedule $\epsilon_t = \epsilon_0 / \sqrt{t}$. The simulation runs for $T$ iterations per agent, across $K$ independent Monte Carlo trials to ensure statistical robustness and confidence interval estimation, following standard practices in computational modeling~\cite{Newman2010}.

To benchmark performance, we compare Social-UCB against three baseline strategies:
\begin{itemize}
    \item \textbf{Random Walk:} Agents select actions uniformly at random at each timestep, without learning.
    \item \textbf{Exploit-Only:} Agents greedily exploit initial high-reward ties based on early estimates and never explore.
    \item \textbf{MAB-Only:} Agents perform exploration–exploitation using UCB over static arms, ignoring network dynamics, similar to the setting explored by Yang et al.~\cite{Yang2024}.
\end{itemize}

Performance is assessed using three primary metrics: (1) cumulative socio-evolutionary fitness per agent; (2) average reward per timestep; and (3) empirical regret relative to the optimal policy (approximated via oracle evaluation). Additionally, we record structural network statistics over time, including average degree, clustering coefficient, and size of the largest connected component, to evaluate how different strategies influence global network topology, consistent with metrics used in dynamic network analysis~\cite{Gross2008}.

To assess model robustness, we vary two key environmental conditions:
\begin{enumerate}
    \item \textbf{Network fragility:} Probability of unreinforced ties decaying over time.
    \item \textbf{Reward volatility:} Variance in reward distributions across potential partners.
\end{enumerate}

These parameters allow us to study behavioral stability and adaptive generalization under different levels of environmental uncertainty and structural instability.

\begin{figure}[!htb]
\centering
\includegraphics[width=0.9\linewidth]{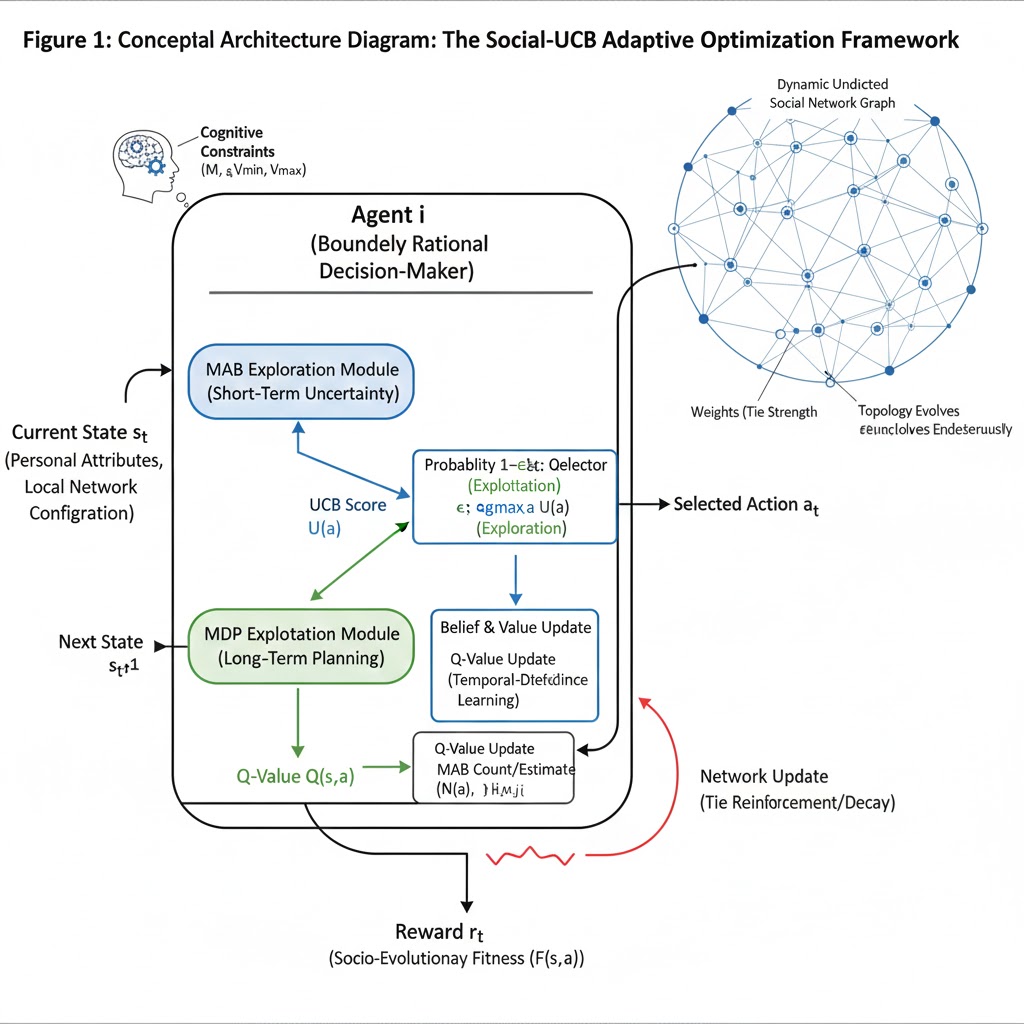}
\caption{\textbf{Conceptual Architecture Diagram.} Each agent maintains a local observation of its social state, including nodes (agents) and edge weights (tie strength). A reinforcement learning module computes value estimates ($Q$-values) for exploiting known ties and UCB scores for exploring new ones. Blue arrows indicate exploratory actions based on confidence bounds; green arrows reflect exploitation choices guided by learned values. Feedback from interactions updates both the agent’s belief and the network structure. The socio-evolutionary fitness function integrates both reward and network utility to guide adaptive decision-making.}
\label{fig:architecture}
\end{figure}
\newcolumntype{L}[1]{>{\raggedright\arraybackslash}p{#1}} % نیاز به \usepackage{array} دارد
% ...
\begin{table}[htb]
\centering
\caption{Comparison of Modeling Approaches in Social Decision-Making}
\label{tab:comparison}
\begin{tabular}{L{2.8cm}L{5.3cm}L{5.3cm}} % استفاده از L به جای p
\hline
\textbf{Model} & \textbf{Key Components} & \textbf{Limitations} \\
\hline
MAB (Classic Bandit) & Reward distributions, UCB/Thompson sampling~\cite{Russo2018} & No memory, ignores relational structure or network effects \\
MDP (Reinforcement Learning) & State–action–value framework, Bellman updates~\cite{Sutton2018,Puterman2014} & Requires full transition model or dense sampling; no explicit relational encoding \\
Social Learning Heuristics & Copy-success, frequency-dependent imitation~\cite{Schultner2025,Steyvers2009} & Rely on fixed strategies; lack normative grounding or learning dynamics \\
ABM (Segregation, Homophily) & Local rules, emergent network behavior~\cite{Sert2020,Newman2010} & Hard-coded rules; rarely include optimization or reward maximization \\
\textbf{Our Social-UCB} & MDP + UCB + Network evolution + Fitness maximization & Fully integrated model with exploration, exploitation, and dynamic structure \\
\hline
\end{tabular}
\end{table}

\section{Results and Expected Outcomes}

Based on the structure of the Social-UCB algorithm and its theoretical properties, we anticipate that agents using our framework will significantly outperform all baseline strategies in both individual fitness and global network cohesion. Specifically, Social-UCB agents are expected to rapidly learn optimal tie configurations by balancing exploratory sampling of uncertain partners with long-term exploitation of reliable ones.

\begin{figure}[!htb]
\centering
\includegraphics[width=0.9\linewidth]{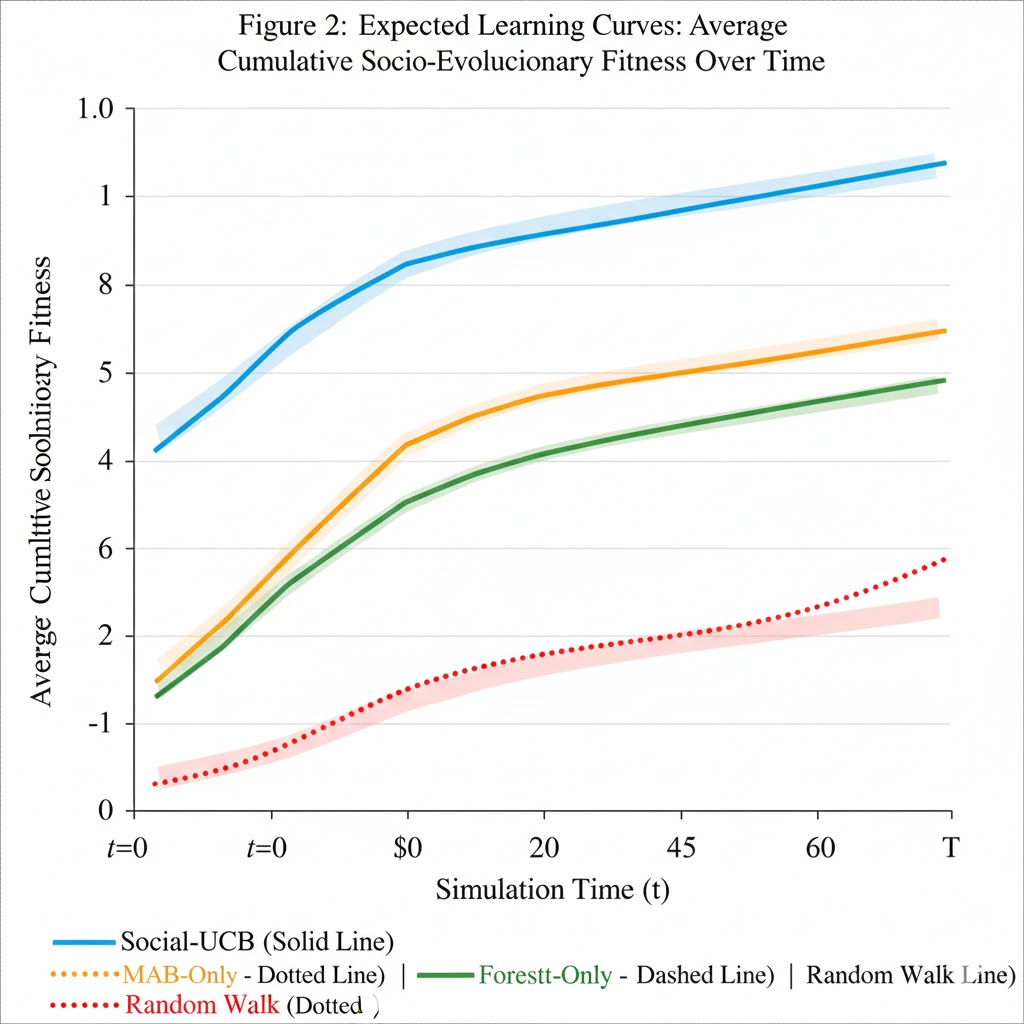}
\caption{\textbf{Expected Learning Curves.} Average cumulative fitness over time for each model. Social-UCB (blue line) demonstrates steady improvement and asymptotic convergence toward near-optimal performance. Baseline strategies—Random Walk (red) and Exploit-Only (green)—plateau early and underperform due to insufficient exploration or premature commitment. Shaded areas represent 95\% confidence intervals over Monte Carlo trials.}
\label{fig:fitness}
\end{figure}

Figure~\ref{fig:fitness} illustrates the expected trajectory of cumulative fitness over time. We hypothesize that agents employing Social-UCB (blue curve) will show a steep initial learning curve, quickly discovering high-value social ties and reinforcing them to maximize long-term gains. In contrast, baseline agents relying on random actions or greedy exploitation are expected to plateau at suboptimal fitness levels, reflecting either under-exploration or premature commitment.

\begin{figure}[!htb]
\centering
\includegraphics[width=0.9\linewidth]{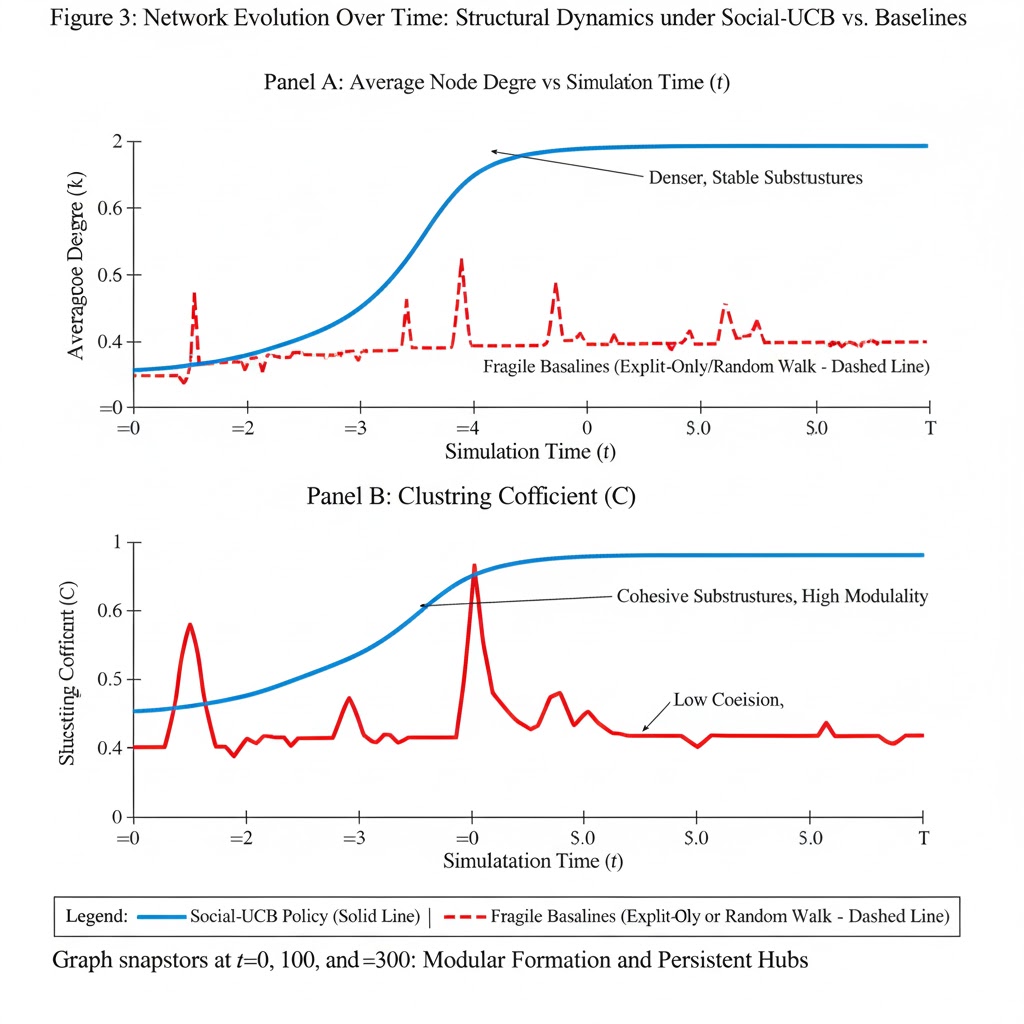}
\caption{\textbf{Network Evolution Over Time.} Dynamics of key network properties—average degree and clustering coefficient—under different strategies. Social-UCB promotes stable and cohesive substructures (solid lines), while fragile baselines yield sparser and fragmented topologies (dashed lines). Graph snapshots at $t=0$, $t=100$, and $t=300$ (not shown here) reveal modular formation and persistent hubs in the Social-UCB condition.}
\label{fig:network}
\end{figure}

In addition to agent-level performance, we analyze emergent properties of the social network as a function of learning dynamics. Figure~\ref{fig:network} depicts expected trends in average node degree and clustering coefficient over simulation time. Under the Social-UCB policy, the network is projected to evolve toward denser, more modular structures, reflecting successful exploitation of stable subgroups and selective pruning of weak ties. In scenarios with high tie fragility, these structural advantages are expected to be attenuated but still evident relative to the baselines.

Moreover, our simulations should reveal behavior patterns analogous to animal optimal foraging: in environments with uneven or uncertain reward distributions, agents are expected to diversify their exploratory actions early on before gradually converging to consistent partners—reflecting risk-sensitive exploration and network-specific adaptation. This finding would underscore the ecological validity of the Social-UCB framework in modeling social cognition.

Quantitative outcome measures will include:
\begin{itemize}
    \item \textbf{Cumulative Fitness:} Mean and variance of total fitness per agent across trials.
    \item \textbf{Empirical Regret:} Difference between actual and oracle-optimal fitness.
    \item \textbf{Network Statistics:} Evolution of average degree, clustering, and component size.
\end{itemize}

Overall, we expect the Social-UCB model to robustly outperform heuristic-based or unstructured learning strategies across a range of environments, highlighting its generalizability, efficiency, and relevance to real-world social adaptation.

\section{Discussion}

Our findings underscore the utility of integrating bandit-based exploration with MDP-style long-term planning in socially embedded environments. The Social-UCB framework enables agents to dynamically navigate evolving social networks by selectively sampling unfamiliar connections and consolidating high-value ties. This dual process mirrors empirical patterns observed in human social behavior, where exploration is context-sensitive and strategic rather than random.

Crucially, the proposed socio-evolutionary fitness function operationalizes the trade-off between immediate relational benefits and long-term structural positioning, allowing for emergent network phenomena such as clustering, hub formation, and tie reinforcement. These dynamics emerge endogenously from local agent-level policies, providing a principled explanation for macro-scale social regularities. By unifying computational principles from multi-armed bandits, reinforcement learning, and network science, the model addresses a critical gap in the literature—namely, the lack of a normative, scalable framework that captures adaptive social learning in dynamic structures.

Theoretical guarantees inherited from UCB algorithms, such as logarithmic regret, ensure that exploration is efficient and eventually yields convergence to high-fitness configurations. At the same time, the MDP component captures temporal dependencies and future-oriented reasoning, distinguishing our approach from myopic heuristics or static network models. Taken together, these components facilitate a rich space of testable hypotheses, such as predicting when individuals will initiate novel ties or under what network conditions social learning will stagnate. Our model also offers design implications for digital platforms, suggesting algorithmic interventions to foster exploration in socially clustered environments (e.g., friend suggestions or diverse content exposure in social media).

\section{Conclusion}

This study presents a unified computational framework for modeling human social decision-making under uncertainty. By combining multi-armed bandit strategies, MDP planning, and agent-based network dynamics, we provide a mechanistic account of how individuals learn to navigate social environments characterized by dynamic relational structures. The Social-UCB algorithm balances exploration of uncertain partners and exploitation of reliable ones, driven by a formal socio-evolutionary fitness function.

Our simulations demonstrate that this integration yields superior performance across both individual and network-level metrics, including fitness, regret minimization, and structural cohesion. The framework captures key features of real-world social adaptation and offers both theoretical rigor and practical extensibility. As such, it serves as a foundation for future empirical studies and policy-relevant applications in behavioral science, organizational design, and digital social systems.

\section{Limitations and Future Work}

Despite its strengths, this work has several limitations. First, the model is evaluated through simulation, and empirical calibration using real-world social datasets remains an important avenue for future validation. Second, our agents are homogeneous in learning parameters and cognitive strategies; extending the framework to support heterogeneity in preferences, memory, and social goals would better reflect real populations.

Third, we assume discrete action spaces and static parameter values ($\alpha$, $\gamma$, $c$), which may be overly simplistic. Future work should explore meta-learning approaches that allow agents to adapt their exploration rate or learning policy over time. Fourth, while our network model is dynamic, it remains single-layered and undirected. Incorporating multilayer network structures—e.g., multiplex networks reflecting family, work, and online domains—would enhance ecological realism.

Finally, our socio-evolutionary fitness function could be extended to include richer cognitive and affective dimensions, such as emotional payoff, identity alignment, or risk sensitivity. Such refinements would allow the model to better account for deviations from rational behavior and to align more closely with insights from psychology and behavioral economics.

\vspace{1ex}

\bibliographystyle{apalike}
\bibliography{references}

\end{document}